\documentclass[aps,prl,amsfonts,superscriptaddress,showpacs,twocolumn]{revtex4}
\usepackage{epsfig}
\usepackage{multirow}
\usepackage{hyperref}
\newcommand{\eq}{\begin{eqnarray}}
\newcommand{\en}{\end{eqnarray}}
\def\bra#1{\mathinner{\langle{#1}|}}
\def\ket#1{\mathinner{|{#1}\rangle}}


\begin{document}

\title{Four-Photon (In)Distinguishability Transition} 
\author{Malte C. Tichy}
\affiliation{Physikalisches Institut, Albert-Ludwigs-Universit\"at, Hermann-Herder-Str.~3, D-79104 Freiburg, Germany}
\author{Hyang-Tag Lim}
\affiliation{Department of Physics, Pohang University of Science and Technology (POSTECH), Pohang, 790-784, Korea}
\author{Young-Sik Ra}
\affiliation{Department of Physics, Pohang University of Science and Technology (POSTECH), Pohang, 790-784, Korea}
\author{Florian Mintert} 
\affiliation{Physikalisches Institut, Albert-Ludwigs-Universit\"at, Hermann-Herder-Str.~3, D-79104 Freiburg, Germany}
\affiliation{Freiburg Institute for Advanced Studies, Albert-Ludwigs-Universit\"at, Albertstr. 19, D-79104 Freiburg, Germany}
\author{Yoon-Ho Kim} 
\affiliation{Department of Physics, Pohang University of Science and Technology (POSTECH), Pohang, 790-784, Korea}
\author{Andreas Buchleitner}
\affiliation{Physikalisches Institut, Albert-Ludwigs-Universit\"at, Hermann-Herder-Str.~3, D-79104 Freiburg, Germany}

\date{\today}

\begin{abstract}
We demonstrate the conspiration of many-particle interferences of different degree to determine the transmission of 
four photons of tunable indistinguishability through a four-port beam splitter array. The probability of
certain output events depends {\em non-monotonically} on the degree of distinguishability, due to distinct
multi-particle interference contributions to the transmission signal.
\end{abstract}
\maketitle

The symmetrization postulate 
imposed by the indistinguishability of particles is a fundamental quantum concept with no classical counterpart, which strongly influences the behavior of matter at any energy scale. 
In quantum optics, the most prominent manifestation thereof is the Hong-Ou-Mandel (HOM) effect \cite{Hong:1987mz}: two indistinguishable photons falling on the opposite input ports of a balanced beam splitter 
always leave the setup together,
{\it i.e.} all amplitudes with two photons in different output modes interfere destructively, while events with both photons in the same mode are enhanced due to the photons' \emph{bosonic nature}. 
When tuning the transition from distinguishability to indistinguishability of the photons, 
which can be described by a \emph{single} parameter \cite{Tichy2}, the visibility of the HOM dip in the probability to detect one photon per mode increases monotonically. 

Signatures for the \emph{full} indistinguishability of {\em more than two} particles can be observed, {\it e.g.}, when many indistinguishable photons bunch at one output mode of a beam splitter \cite{Ou:2008zv,Xiang:2006tw,Niu:2009pr}. In addition, when several particles enter a multiport beam splitter \cite{PhysRevA.71.013809} simultaneously, \emph{many-particle interferences} 
strongly influence the probability of individual counting events, 
such that many distinct events with a given number of particles per output port are totally suppressed \cite{Campos:2000yf,Lim:2005qt,Tichy:2010kx}. 
The \emph{transition} between fully distinguishable and fully indistinguishable particles has not received attention beyond its impact on bosonic bunching \cite{Xiang:2006tw}. Its role for the behavior of other events, {\it e.g.} the ones that are suppressed for indistinguishable particles \cite{Lim:2005qt,Tichy:2010kx}, is widely open. A thorough understanding of partial indistinguishability is, however, mandatory for the experimental characterization of the degree and nature of many-particle interferences. 

In the present Letter, we 
consider four photons that propagate through a four-port beam splitter array.
In contrast to an intuitive extrapolation of the 
features of the well-known two-photon case \cite{Hong:1987mz}, and of the bunching behavior of many photons \cite{Xiang:2006tw}, we show that (i) the degree of distinguishability manifests itself in 
{\em non-monotonic} 
event probabilities, and that (ii) events with large occupation numbers are not necessarily more 
likely with increasing indistinguishability,
despite the bosonic nature of photons. 
These results are established by quantitative predictions on experimentally directly accessible quantities. 

\begin{figure}[h]
\includegraphics[width=7.5cm,angle=0]{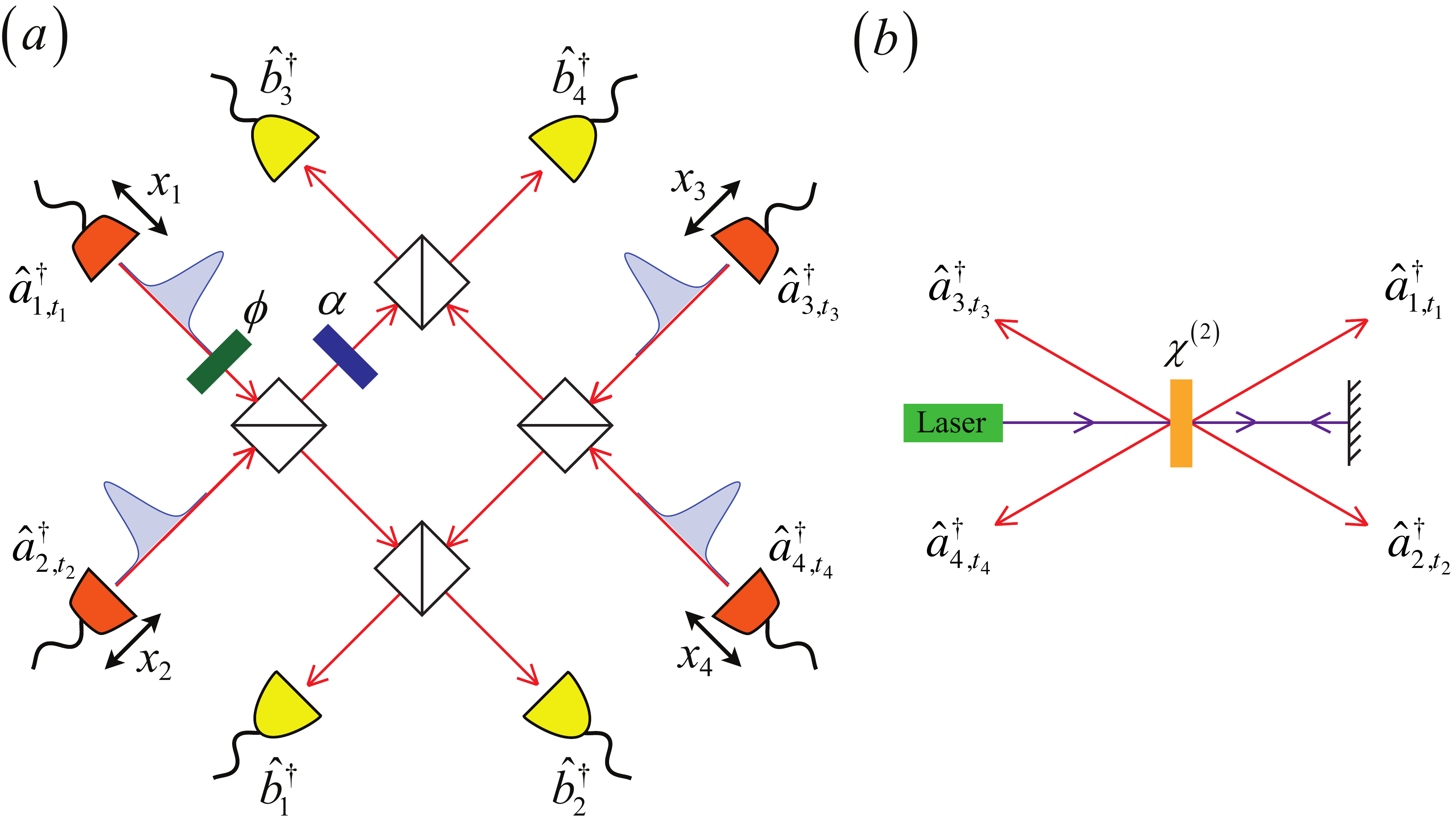}
\caption{(color online) (a) Diamond-shaped multiport (balanced) beam splitter array. Four ingoing modes $\hat a^\dagger_i$ are redistributed onto the output modes $\hat b^\dagger_j$. The path length $x_j$ of each incoming mode controls the mutual distinguishability of the particles in the setup. (b) Creation of four photons by spontaneous parametric down conversion, by double-passage of a laser pulse through a non-linear crystal.}\label{setup}
\end{figure}

We study many-particle interference within a setup with four ports \cite{PhysRevA.55.2564}, see Fig.~\ref{setup}(a). In this arrangement, no single-particle (Mach-Zender-like) interference can occur, what makes it
perfect to study the genuine manifestation of many-particle interference. 

Single-particle evolution is here described by a unitary matrix which relates particle creation operators of input and output ports, $\hat a_{i,\omega}^\dagger$ and $\hat b_{k,\omega}^\dagger$, respectively, via \eq 
\hat a_{i,\omega}^{\dagger} \rightarrow \sum_{k=1}^4 U_{ik}(\alpha,\phi) \hat b_{k,\omega}^\dagger, \en  with the photon frequency $\omega$ unchanged, and $U(\alpha,\phi)$ given by
\eq U(\alpha, \phi) &=& \frac 1 2 \left( \begin{tabular}{cccc}
$e^{i \phi}$ & $ e^{i \phi}$ & $e^{i (\phi+\alpha)} $&$e^{i (\phi+\alpha)}$ \\
1 & 1 & $-e^{i \alpha} $ &$-e^{i \alpha} $ \\
1 & $-1 $& 1 & $-1$ \\
1 & $-1 $& $-1$ &1 \\  
\end{tabular} \right)\ 
\label{matrix} .\en
For any value of $\alpha$ and $\phi$, the matrix $U(\alpha, \phi)$ is a complex unitary 
matrix with $|U_{jk}|=1/2$, {\it i.e.} a Hadamard matrix \cite{Tadej:2006it}.  The phases $\alpha, \phi$ have 
two rather distinct physical interpretations: $\phi$ absorbs all relative phases between input modes, and is well-known from the two-mode interference of many particles, {\it e.g.} in N00N-state interferometry \cite{Afek}. $\alpha$ is of distinct origin: It corresponds to the phase enclosed by the interfering modes (similar to a Sagnac interferometer \cite{sagnac}), emerges only in the four-mode case \cite{Tadej:2006it}, and effectively controls the relative phases of the \emph{output} components, conditioned on the input modes. 

Four photon states are created by the double passage of a laser pulse through a non-linear crystal, through spontaneous parametric down conversion, see Fig.~\ref{setup}(b). Such states consist of the coherent superposition of a \emph{quadruplet} part, with one photon in each mode, created when the laser pulse induces one pair of photons at each time it passes the crystal, with two \emph{double-twin} parts, where 
four photons are distributed among two modes,  resulting from events where the pulse generates two pairs in the one or in the other passing direction \footnote{The relative phase between double-twin and quadruplet contribution is absorbed in the phase $\phi$, eq.~(\ref{matrix}).}. Since all three processes occur with the same probability, the initial state reads
 \eq \ket{\Psi}=\frac{1}{\sqrt{3}} \left( \prod_{j=1}^4 \hat a^\dagger_{j,t_j} 
 +
  \frac { \hat a_{1,t_1}^{\dagger~2} \hat a_{2,t_2}^{\dagger~2}}{2} +  
   \frac {\hat
a_{3,t_3}^{\dagger~2} \hat a_{4,t_4}^{\dagger~2}}{2}   \right)\ket{0} \label{inistate} .
\en
The operator
\eq \hat a_{j,t_j}^\dagger =
\int_{-\infty}^{\infty} \mbox{d} \omega \frac{1}{\sqrt \pi
\Delta \omega} e^{- \frac{(\omega-\omega_0)^2}{2 \Delta\omega^2}
} e^{i \omega t_j} \hat a_{j,\omega}^\dagger \en
creates a single photon with central frequency $\omega_0$ and spectral width $\Delta \omega$ at the input port $j$, at time $t_j$. The arrival times $t_j$ of the photons can be tuned through variable path lengths $x_j=c \cdot t_j$ depicted in Fig.~\ref{setup}(a), and hence their
overlap, or indistinguishability, by virtue of
$ |\bra 0 \hat a_{i,t_j} \hat a^\dagger_{i,t_k} \ket 0|^2 
=\text{exp}({-\frac 1 2 \Delta\omega^2 (t_j-t_k)^2})$ 
A change of relative path lengths induces a phase-shift between the input-modes, which can be accounted for in $\phi$:
\eq \phi \rightarrow  \phi + \omega(t_1+t_2-t_3-t_4) \label{thephi}  \en 

All possible final states that emerge from the multiport can be characterized in terms of the photon number 
detected in each port. For four photons, simple combinatorics yields 35 distinct events which can be labelled with vectors $\vec s=(s_1,s_2,s_3,s_4)$, where $s_j$ is the number of particles in port $j$. Their 
order is hereafter given by their relative abundance in the fully distinguishable case, eq.~(\ref{combi}) below, and such that vectors $\vec s$ with large $s_1$ come first, i.e., $\vec s_1=(4,0,0,0), \vec
s_2=(0,4,0,0), \dots, \vec s_{5}=(3,1,0,0),$ and, finally, $\vec s_{35}=(1,1,1,1)$.

For the interpretation of the interference effects that we will discuss hereafter, it is useful to decompose the initial state (\ref{inistate}) into a superposition of states in which the photons in different modes are all either fully indistinguishable or fully distinguishable. In the two-photon case, only two contributions 
(distinguishable and indistinguishable) arise. For four photons, the Gram-Schmidt-orthogonalization of the single-particle states that appear in (\ref{inistate}) yields $4!=24$ different terms for the four-particle state, since the description of the $j$th particle needs up to $j$ non-vanishing components. Each term corresponds to a certain \emph{distinguishability setting} that we will denote by $\{i_1,i_2,i_3,i_4\}$, where photons in port $k$ and $l$ are indistinguishable if $i_k=i_l$. Fully 
indistinguishable/distinguishable particles correspond to $\{1,1,1,1\}$/$\{1,2,3,4\}$, and only indistinguishable particles interfere. Hence, the distinguishability
setting determines the degree of many-particle interference.
In general, a setup with given arrival times for all photons 
corresponds to a situation in which several distinguishability settings contribute to the initial state, with different degrees of multi-particle interference occurring simultaneously.

The extreme case of full distinguishability $\{1,2,3,4\}$ is realized when all photons have pairwise delays 
$|t_i-t_j|\gg 1/ {\Delta \omega}$. In this case, we can safely neglect the exponentially suppressed components that still exhibit many-particle interference, and simple 
combinatorics can be applied to yield the output event probabilities (remember that no single-particle interference can occur in our setting): 
\eq P_{\text{dist}}(\vec s)=\frac{4!}{4^4 \prod_j s_j!} \, . \label{combi} \en 
The opposite, fully indistinguishable limit  $\{1,1,1,1\}$ is realized for $t_1=t_2=t_3=t_4$, when all photons can interfere perfectly. 

\begin{figure}[h] \center
 \includegraphics[width=7.75cm,angle=0]{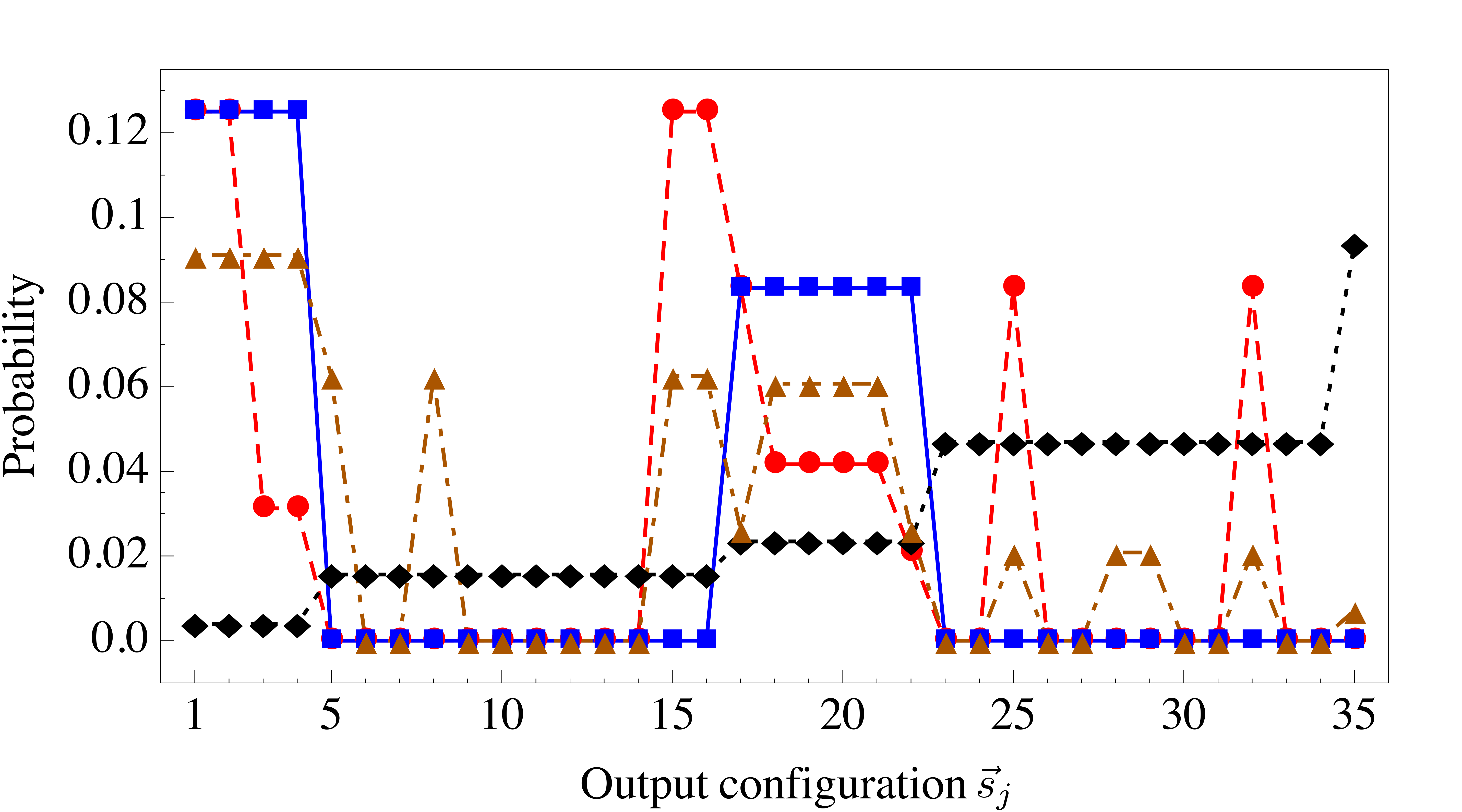}
\caption{(color online) Event probabilities of all 35 possible output configurations $\vec s_j$, for distinguishable (black diamonds) and fully indistinguishable particles, with $\phi=\alpha=0$ (blue squares), $\phi=0,\alpha=\pi/4$ (red circles) and $\phi=\pi/4, \alpha=0$ (brown triangles).}\label{quantClass1}
\end{figure}

The resulting probabilities for five representative events are compared in Tab.~\ref{tab1}, for the fully distinguishable and indistinguishable case, respectively. The event $\vec s_{14}=(0,1,0,3)$ is fully suppressed for indistinguishable photons, for any choice of the 
phases $\alpha, \phi$, while $\vec s_{21}, \vec s_{22}$ and $\vec s_{35}$ exhibit an intricate $\alpha,\phi$-dependence. Most importantly, there is no unambiguous correlation between the event probability and the photon distribution on the output modes: as anticipated above, $s_1=(4,0,0,0)$ may be enhanced (as expected for bosonic bunching) as well as strictly suppressed (for the experimentalist's choice $\phi=\pi$). Unexpectedly, $\vec s_{35}=(1,1,1,1)$ may be enhanced up to a weight $32/9$. 
\begin{table}[h]
\begin{tabular}{r|r|r} 
Event  & $P_{\text{dist}}$ \vspace{0.1cm} & $P_{\text{id}}(\vec s)$ \\ \hline  
$\vec s_{1}=(4,0,0,0)$ & 1/256 & $\cos^4( \phi/2)/8 $\\
$\vec s_{14}=(0,1,0,3)$ & 1/64 & 0 \\
$\vec s_{21}=(0,2,0,2)$ & 3/128 & $\left[\cos (\alpha) + \cos(\alpha+\phi)\right]^2/48$\\
$\vec s_{22}=(0,0,2,2)$ & 3/128 & $\left[1 - 3 \cos(2 \alpha+\phi)\right]^2/48$\\
$\vec s_{35}=(1,1,1,1)$ & 3/32 & $\left[\cos (\alpha) - \cos(\alpha+\phi)\right]^2/12$\\
\end{tabular}
\caption{Event probabilities for fully distinguishable ($P_{\text{dist}}$) and indistinguishable particles ($P_{\text{id}}$). }
\label{tab1}
\end{table}
In stark contrast to the two-photon HOM effect, the manifestation of perfect indistinguishability is not unique in the present, multi-particle interference scenario, and has no intuitive interpretation in terms of the occupation of modes. This is highlighted by Fig. \ref{quantClass1}, which displays all event probabilities, for fully distinguishable and indistinguishable particles, 
and three different choices of $\alpha$ and $\phi$, and thus demonstrates the loss of any indistinguishability-induced hierarchy in the event probabilities. Furthermore, not only are the event probabilities no unambiguous witnesses of (in)distinguishability any more, but they even evolve {\em non-monotonically} with decreasing distinguishability of the particles, as can be demonstrated in our setting, by subsequently rendering different pairs of photons indistinguishable, and thus adding interference terms between an increasing number of photons. 
\begin{figure*}[t]
\includegraphics[width=\linewidth,angle=0]{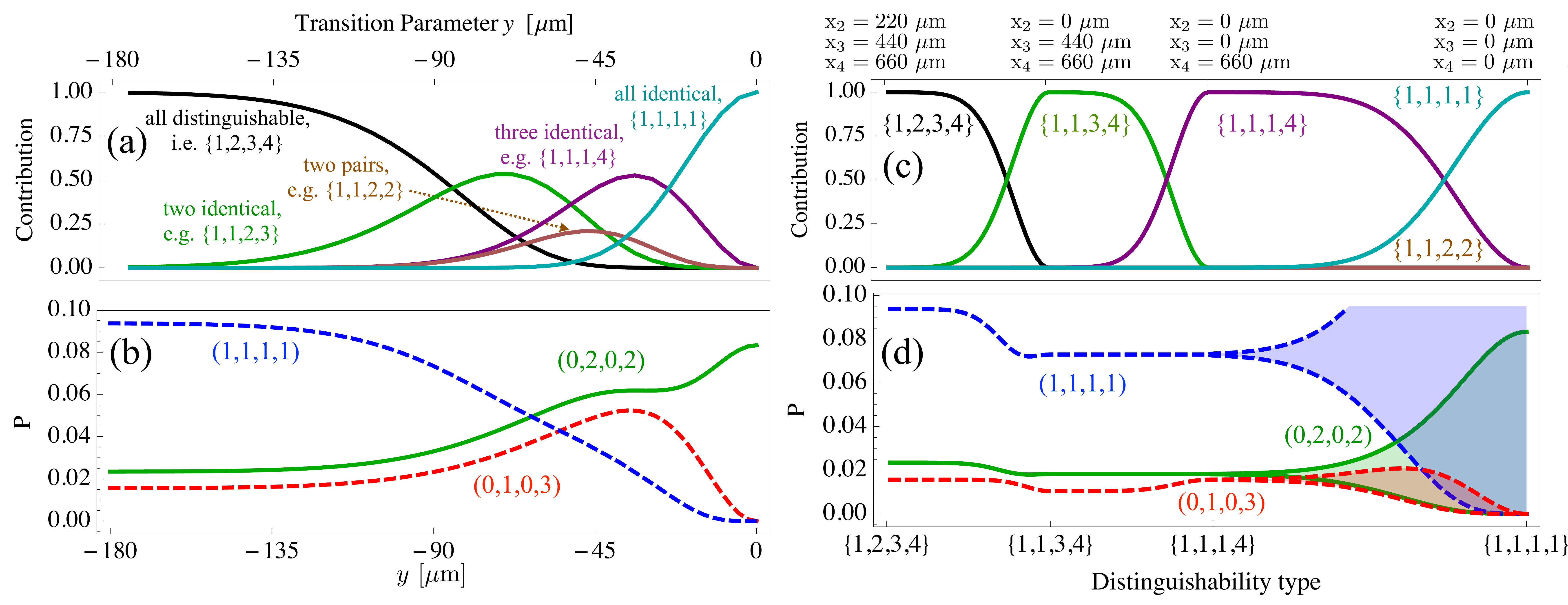}
\caption{(color online) Evolution of event probabilities during the transition from distinguishable to indistinguishable particles (bottom row). Left: continuous transition, 
parametrized by $y$, with $\alpha=0$. Right: step-wise transition, with corresponding interferometer path lengths, Fig.~\ref{setup}, on top of 
panel (c), with $x_1=0\ \rm \mu m$, $\alpha=\pi$. 
(a,c): Distinguishability type contributions to the output signal: While for the step-wise transition, at most two distinguishability settings are relevant at a time, several may contribute in 
the continuous transition. (b,d): Event probabilities of $\vec s_{35}=(1,1,1,1)$ (blue), $\vec s_{14}=(0,1,0,3)$ (red) and $\vec s_{21}=(0,2,0,2)$ (green). The shaded areas in (d) 
represent four-photon interference fringes oscillating as a function of $\phi$. 
 }\label{Trans1}
\end{figure*}

Let us focus on the effects of partial interference on the three events $\vec s_{35}=(1,1,1,1), \vec s_{14}=(0,1,0,3), \vec s_{21}=(0,2,0,2)$ (all listed in Tab.~\ref{tab1}), within two exemplary distinguishability transitions: For specificity, we assume that the single photons have a central wavelength of $\lambda_0=780$~nm, and a full width at half maximum (FWHM) of $\Delta_\lambda=5$~nm, what corresponds to a coherence length of $l_c\approx$~122 $\mu$m. 

\emph{(I)} We first adjust the path lengths in a continuous way, by parametrizing $x_1=0$, $x_2=y$, $x_3=-y$ and $x_4=2 y$. Hence, for $y=-180~\mu$m, the particles are fully distinguishable $\{1,2,3,4\}$, while for $y=0~\mu$m, their wave functions fully overlap, the particles are fully indistinguishable $\{1,1,1,1\}$. During the transition, several distinct contributions to the output signal arise simultaneously (see Fig.~\ref{Trans1}a). We initially choose $\alpha=0$ and $\phi=0$. The latter parameter remains constant for all values of $y$ with our chosen parametrization (cf. (\ref{thephi})). Consequently, only variations on length-scales of the order of the coherence length $l_c$ appear. 

The signal evolution during the (controlled) indistinguishability transition is shown on the bottom left
of Fig.~\ref{Trans1}: The event $\vec s_{35}=(1,1,1,1)$ is progressively suppressed as higher order interference contributions, from two- over three- to four-photons, kick in. This contrasts with the clearly non-monotonic event probability of $\vec s_{14}=(0,1,0,3)$. While two- and three-photon-interference increase its probability, it is eventually strictly suppressed for fully indistinguishable particles, resulting in a 
maximum for partial distinguishability. For $\vec s_{21}=(0,2,0,2)$, a step-like-behavior  results from the competition of destructive three-particle- and constructive four-particle interference. 

\emph{(II)} Another remarkable impact of the distinguishability transition is born out when we adjust the path lengths step-by-step: Starting from fully distinguishable photons, with $x_1=0$, $x_2=220~\mu $m, $x_3=440~\mu$m, $x_4=660~\mu$m,  we first tune down $x_2$, and subsequently $x_3$ and $x_4$, as indicated in Fig.~\ref{Trans1}(c), with $\alpha =\pi$. The thus chosen, dominant distinguishability settings are $  \{1,2,3,4 \} \rightarrow \{1,1,3,4 \} \rightarrow \{1,1,1,4 \} \rightarrow \{1,1,1,1 \}$, and this step-wise 
transition is reflected by the event probabilities in Fig.~\ref{Trans1}(d). 
In particular, the probability of $\vec s_{14}=(0,1,0,3)$ is again non-monotonic, due to a {\em destructive} two-photon contribution, followed by a {\em constructive} three-photon 
contribution for the $\{1,1,1,4\}$ setting.

An additional effect manifests when four-photon interference sets in: the event probabilities start to depend 
on the phase $\phi$. This leads to fast oscillations on the scale of the photon wavelength $\lambda$, an intrinsic feature of the interference of four or more photons, shown in the plots as shaded area between 
minima and maxima. The dependence of the event probability on $\phi$ for fully indistinguishable photons (Tab.~\ref{tab1}) explains the onset of fast oscillations of the event probabilities for $\vec s_{21}$ and $\vec s_{35}$, and the growth of their amplitudes with increasing $\{1,1,1,1\}$ contribution. Unexpectedly, also $\vec s_{14}=(0,1,0,3)$ exhibits such oscillations which, however, vanish when four-photon interference fully dominates, since this event is then strictly suppressed (Tab.~\ref{tab1}). It is the interplay of the $\{1,1,1,4\}$ and $\{1,1,1,1\}$ settings that leads to such dependence on $\phi$: the double-twin 
part of the wave function in the setting $\{1,1,1,4\}$ interferes with the quadruplet part in the $\{1,1,1,1\}$ setting, resulting in four-photon interference that depends on $\phi$, and a local maximum of the event probability at the point where the contributions of $\{1,1,1,4\}$ and $\{1,1,1,1\}$ are equal. This feature is specific to the \emph{partial} distinguishability of the photons. 

The discussed phenomena constitute show-case examples for the intricate effects that manifest when quantifying or controlling the particles' degree of indistinguishability in an experiment. Despite the mere doubling of the number of particles with respect to the HOM setup, the (in)distinguishability transition cannot be explained any more by extrapolation of the two-photon effect: interference dominates over bosonic bunching, and events with large occupation numbers are not necessarily enhanced when approaching indistinguishability. The phase $\alpha$ which quantifies the phase enclosed  by the setup constitutes an input- and output-mode dependent parameter 
with measurable impact, as a physical consequence emerging from the formal definition of four-dimensional complex Hadamard matrices \cite{Tadej:2006it}. Finally, not only the degree, but also the quality of interference changes with the number of interfering particles. 

Our results impressively demonstrate that we need to fully abandon the idea that many-particle interference manifests itself in a unique fashion that 
can be predicted from the bosonic nature of particles alone, as usually stated in the two-particle case \cite{jeltes}. Constructive and destructive interference effects occur for \emph{all} final 
events, and they can be turned into one another by variation of the phases.
Our decomposition of the initial state into distinguishability settings does not only offer a powerful tool for the computation of event probabilities, necessary for the experimental implementation 
of many-particle interference, but also allows us to understand and classify the occurring phenomena.
Due to the combinatorial explosion of possible many-particle paths, a description of the indistinguishability transition becomes prohibitive for many more than four particles, and it remains to be studied whether one can find other, coarse-grained observables \cite{Mayer} with a monotonic dependence on a single parameter, to quantify many-particle distinguishability.

We thank Jaewan Kim and Joonwoo Bae for fruitful discussions and hospitality, during the KIAS Workshop on Quantum Information Sciences 2009, 
where this work was initiated. Financial support by the DAAD-GEnKO programme (M.C.T., F.M., A.B.), by the National Research Foundation of Korea 
(2009-0070668 and KRF-2009-614-C00001; H.-T.L., Y.-S.R., Y.-H.K.), by the German Academic National Foundation (M.C.T.), and through DFG grant MI 1345/2-1 (F.M.) is gratefully acknowledged.

\end{document}